\documentclass{article}
\newlength{\abstractwidth}
\usepackage{cancel}
\usepackage{hyperref}
\usepackage{color}
\usepackage{graphicx}
\usepackage{tikz}

\usepackage{amsmath}
\usepackage{amssymb}
\usepackage{latexsym}
 \usepackage{setspace}
 
 \usepackage{caption}
  \usepackage{cite}
\usepackage{subcaption}

\numberwithin{equation}{section}

\renewcommand{\thefootnote}{\fnsymbol{footnote}}
\renewcommand{\thanks}[1]{\footnote{#1}}
\newcommand{\starttext}{
\setcounter{footnote}{0}
\renewcommand{\thefootnote}{\arabic{footnote}}}
\newcommand{\bea}{\begin{eqnarray}}
\newcommand{\eea}{\end{eqnarray}}
\newcommand{\be}{\begin{eqnarray}}
\newcommand{\ee}{\end{eqnarray}}

\def\ie{\begin{equation}\begin{aligned}}
\def\fe{\end{aligned}\end{equation}}

\def\ie{\begin{equation}\begin{aligned}}
\def\fe{\end{aligned}\end{equation}}

\def\s{\sigma}

\begin{document}

\starttext

\setcounter{footnote}{0}

\begin{flushright}
%\scriptsize 
{\small QMUL-PH-21-30}
\end{flushright}

\vskip 0.3in

\begin{center}

{\large \bf Notes on gravity multiplet correlators in $AdS_3 \times S^3$}

\vskip 0.2in

{Congkao Wen, and Shun-Qing Zhang} 
   
\vskip 0.15in

{\small  School of Physics and Astronomy, Queen Mary University of London, }\\ 
{\small  London, E1 4NS, UK}

\vskip 0.15in

{\tt \small c.wen@qmul.ac.uk, shun-qing.zhang@qmul.ac.uk}

\vskip 0.5in

\begin{abstract}
\vskip 0.1in

We present a compact formula in Mellin space for the four-point tree-level holographic correlators of chiral primary operators of arbitrary conformal weights in $(2,0)$ supergravity on $AdS_3 \times S^3$, with two operators in tensor multiplet and the other two in gravity multiplet. This is achieved by solving the recursion relation arising from a hidden six-dimensional conformal symmetry.  We note the compact expression is obtained after carefully analysing the analytic structures of the correlators. Various limits of the correlators are studied, including the maximally R-symmetry violating limit and flat-space limit. 

   \end{abstract}                                            
   
\end{center}

\baselineskip=15pt
\setcounter{footnote}{0}

\newpage

\setcounter{page}{1}
\tableofcontents

\newpage

%%%%%
\section{Introduction}
\label{secintro}
 %%%%%

Over past several years, many powerful methods have been developed to study the correlation functions of local operators in holographic CFTs when it is possible to exploit the dual description in terms of a weakly coupled gravity theory.  A beautiful example is the study of holographic correlators in type IIB string theory on $AdS_5 \times S^5$. In particular, in supergravity limit, a very compact formula in Mellin space was obtained for the tree-level correlation functions of four one-half BPS single-trace operators with arbitrary conformal weights \cite{Rastelli:2016nze,Rastelli:2017udc}. These results, combining with other powerful techniques and ideas, such as analytical bootstrap program, supersymmetric localisation, have led to many further developments in understanding the holographic correlation functions in $AdS_5 \times S^5$. For instance, the methodology allows constructing loop corrections to these correlators as well as determining higher-derivative contributions arising from $\alpha'$-expansion of superstring  theory \cite{Alday:2017xua, Aprile:2017bgs, Aprile:2017xsp, Aprile:2017qoy, Alday:2017vkk, Aprile:2018efk, Alday:2018pdi, Aprile:2019rep, Chester:2019pvm, Alday:2019nin, Drummond:2019hel, Drummond:2020uni, Aprile:2020luw, Drummond:2020dwr, Bissi:2020wtv, Bissi:2020woe, Binder:2019jwn, Chester:2019jas, Chester:2020dja, Chester:2020vyz, Green:2020eyj, Dorigoni:2021bvj, Dorigoni:2021guq}.

One of the remarkable properties of these holographic correlators in $AdS_5 \times S^5$ supergravity, which will be mostly relevant for our study, is that the correlators of operators with different conformal weights are in fact all related to each other due to a hidden ten-dimensional conformal symmetry \cite{Caron-Huot:2018kta}. This hidden conformal symmetry allows packaging the four-point holographic correlators of operators with different conformal weights into a single four-point correlator of scalar operators in ten dimensions. This observation leads to a recursion relation that relates holographic correlators of operators with higher weights to the lowest-weight one, and the solution of the recursion relation yields precisely the formula originally given in \cite{Rastelli:2016nze}. This observation has been further explored for the holographic correlators in $AdS_5 \times S^5$ beyond the tree-level approximation~\cite{Alday:2019nin}, and the fate of the hidden symmetry, when higher-derivative stringy effects are included, has been studied in~\cite{Drummond:2019odu, Abl:2020dbx, Aprile:2020mus}. 

Analogous hidden conformal symmetries have been found for other holographic theories. In particular, it was found that four-point tree-level holographic correlators of type IIB supergravity in $AdS_3 \times S^3 \times M_4$ (the internal space $M_4$ can be $T^4$ or $K3$ surface) also exhibit a hidden symmetry \cite{Rastelli:2019gtj, Giusto:2020neo} \footnote{See recent work \cite{Alday:2021odx} for the study of hidden symmetries of correlation functions in other holographic theories.}. In the case of $AdS_3 \times S^3$, the hidden symmetry was found to be a six-dimensional conformal symmetry. Supergravity in $AdS_3 \times S^3$ has half maximal supersymmetry, and contains two super multiplets: gravity multiplet and tensor multiplet. It was found that for the correlators in tensor multiplet, just as in the case of $AdS_5 \times S^5$, the tree-level holographic correlators of four chiral primary operators (CPOs) in tensor multiplet can again be packaged into a single correlator of scalar operators in a six-dimensional CFT. This six-dimensional correlator serves as a generating function, which generates holographic correlators of operators with arbitrary weights. Equivalently, the hidden symmetry leads to a recursion relation which can be solved explicitly and gives rise to a compact formula for all tree-level four-point holographic correlators in $AdS_3 \times S^3$ for CPOs in tensor multiplet, which puts our understanding of these correlators at the same level of the holographic correlators in $AdS_5 \times S^5$. 

The hidden six-dimensional conformal symmetry for holographic correlators in $AdS_3 \times S^3$ has recently been extended to more general four-point correlators in~\cite{Giusto:2020neo}, including those with CPOs in gravity multiplet. It was found that the correlators of operators in gravity multiplet are described by a single correlator involving self-dual $3$-forms (instead of simple scalars as in the case of tensor multiplet) in a six-dimensional CFT. This paper aims at a better understanding of holographic correlators in $AdS_3 \times S^3$, especially with operators in gravity multiplet. We will be mainly concerned with the mixed correlators, with two operators in gravity multiplet and the other two in tensor multiplet. 

With this new understanding of the hidden conformal symmetry, again a recursion relation was obtained for the mixed correlators of operators in both tensor and gravity multiplets~\cite{Giusto:2020neo}. The main focus of this paper is to solve the recursion relation, and to obtain a compact formula for the mixed correlators of operators with arbitrary conformal weights. The results arising directly from solving the recursion relation are in fact rather lengthy and complex for the correlators involving operators in gravity multiplet. They are greatly simplified by a better understanding of their analytic structures in Mellin space. We will also make contact with the known flat-space scattering amplitudes by taking flat-space limit on the Mellin amplitudes, and find a perfect match. 

The paper is organised as follows.  In section \ref{sec:hidden-tensor}, we will begin by reviewing the hidden 6D conformal symmetry for the holographic correlators in $AdS_3 \times S^3$ of operators in tensor multiplet as well as the recursion relation arsing from the hidden conformal symmetry. We will solve the recursion relation and obtain a compact formula for the correlators in Mellin space. The study of correlators in tensor multiplet serves a warm-up for understanding the correlators with operators in gravity multiplet. In this section \ref{secgravity}, we will review the main results~\cite{Giusto:2020neo}, where a recursion relation was obtained from the hidden 6D conformal symmetry for mixed correlators involving operators in both tensor and gravity multiplets. However, we find that the solution arising from solving directly the recursion relation is rather lengthy and it is obscured to any simple structures in the formula. The result can be greatly simplified by carefully analysing the analytic structures of the correlators in Mellin space, and a much more compact expression is obtained. We verify the final expression is local, in the sense that all the multiple poles cancel out non-trivially. We further study various limits (including flat-space limit and maximally R-symmetry violating limit) of the results. We conclude and comment on future research directions in section \ref{secconclusion}.

 %%%%%%
  \section{Hidden 6D conformal symmetry in $AdS_3 \times S^3$}
\label{sec:hidden-tensor}
%%%%%%

It was observed in \cite{Caron-Huot:2018kta} that tree-level holographic correlation functions of four BPS scalar operators in $AdS_5 \times S^5$ obeys a remarkable 10D hidden conformal symmetry, which allows packaging the correlators of operators with different conformal weights into a single correlator of four scalars in a 10D CFT.  This 10D hidden conformal symmetry leads to a powerful recursion relation that relates correlators of operators with higher weights to those with lowest weights. Solving the recursion relation explicitly gives rise to a compact expression for all tree-level correlators of four BPS operators, in agreement with the results of \cite{Rastelli:2016nze}.

This observation has been extended to four-point correlation functions in $AdS_3 \times S^3 \times M_4$, where the internal manifold $M_4$ can either be four torus $T^4$ or the $K3$ surface. Compared to the case of $AdS_5 \times S^5$, due to the fact that the $AdS_3 \times S^3$ background has only half maximal supersymmetry, the structure of hidden symmetry is more involved. In particular, the 6D $(2,0)$ supergravity contains tensor and gravity multiplets, for the correlators of operators in tensor multiplet, the situation is very similar to the case of $AdS_5 \times S^5$, where the four-point tree-level correlators of BPS operators with different conformal weights are packaged into a single four-point correlation function of scalars in a 6D CFT \cite{Rastelli:2019gtj}. However, when the operators in  gravity multiplet are involved, the structures become much richer. It was understood in \cite{Giusto:2020neo} that the tree-level four-point holographic correlators with two operators in tensor multiplet and two in  gravity multiplet are, again, described by a single CFT$_6$ correlator, but now with two scalars and two self-dual $3$-forms in six dimensions. 

Below we will review the hidden 6D conformal symmetry, especially its implications on the recursion relation for the four-point holographic correlators in $AdS_3 \times S^3$. As a warm-up, we will begin with the simpler case where all the operators are in tensor multiplet. In next section, we will study the more involved case, where the correlators contain operators with two of them in tensor multiplet and the other two in gravity multiplet. 

 %%%%%%
  \subsection{Four-point correlators of operators in tensor multiplet}
\label{sec:tensor}
%%%%%%

\subsubsection{Flat space superamplitudes}

In flat-space, the four-point superamplitude in 6D $(2,0)$ supergravity of states in tensor multiplet takes a simple form, given as \footnote{More details regarding all tree-level superamplitudes in 6D $(2,0)$ supergravity in flat-space can be found in \cite{Heydeman:2018dje, Schwarz:2019aat}.}
\ie \label{eq:flat}
\mathcal{A}_4^{\rm tensor} = G_6\, \delta^8 \left(\sum_{i=1}^4 q_i \right) \delta^6 \left(\sum_{i=1}^4 p_i \right) \left({\delta_{f_1 f_2} \delta_{f_3 f_4} \over \bold{s}} + {\delta_{f_2 f_3} \delta_{f_1 f_4} \over \bold{t}}+ {\delta_{f_1 f_3} \delta_{f_2 f_4} \over \bold{u}} \right) \, ,
\fe
where $\bold{s}, \bold{t}, \bold{u}$ are Mandelstam variables defined as $\bold{s}=(p_1+p_2)^2, \bold{t}=(p_2+p_3)^2, \bold{u}=(p_1+p_3)^2$, obeying $\bold{s} +\bold{t}+ \bold{u}=0$. And $G_6$ is the Newton constant in 6D and $\delta^8(\sum_{i=1}^4 q_i)$ is due to the conservation of the supercharge, which reflects 6D $(2,0)$ supersymmetry. Explicitly, the supercharge is defined as $q^{A, I}_i= \lambda^A_{i, a} \eta^{a, I}_i$, with $I=1,2$. Here we have used the spinor-helicity formalism for 6D massless momentum 
\ie
p_{i\,\mu} (\Gamma^{\mu})^{AB} = \lambda^A_{i\,a} \lambda_{i}^{B, a} =  {1\over 2} \epsilon^{ABCD}  \tilde{\lambda}_{i\,, C\, \hat{a}} \tilde{\lambda}_{i\, D}^{\hat a} \, .
\fe    
The index $i$ indicates the external particle, $\mu$ is a vector index and $A,B,\ldots,$ are spinor indices of 6D Lorentz group, and $a$ and $\hat{a}$ are $SU(2) \times SU(2)$ indices labelling the 6D little group $SO(4)$ for massless particles. To describe the 6D $(2,0)$ supersymmetry we have introduced Grassmann variables $\eta^{a, I}_i$, which can be used to package all the on-shell states into a on-shell superfield \cite{Heydeman:2018dje}. Finally, $f_i$ are flavour indices of the states in tensor multiplet. 

After stripping off the supercharge conservation factor $\delta^8(\sum_{i=1}^4 q_i)$, the amplitude is identical to the four-point tree-level amplitude of $\phi^3$ theory (with a flavour symmetry), which is conformal invariant in 6D. This property (and the results of $AdS_5 \times S^5$) has led to the conjecture of hidden 6D conformal symmetry for four-point tree-level holographic correlators in $AdS_3 \times S^3$ of operators in tensor multiplet.

\subsubsection{Hidden 6D conformal symmetry and recursion relation in tensor multiplet}
\label{sec:tensor-cor}

In general, four-point correlators of CPOs in the 2D CFT that is dual to type IIB string theory on $AdS_3 \times S^3$ that we will consider take the following form, 
\ie \label{eq:def}
\langle O_{k_1} O_{k_2} O_{k_3} O_{k_4}    \rangle =  \left( \frac{|\zeta_{13}|^{k^-_{21} + k_{43}^-} |\zeta_{23}|^{-k^-_{21} + k_{43}^-}}{|\zeta_{12}|^{k^+_{12} + k_{43}^-} |\zeta_{34}|^{2k_4}}  \right) \left[  \mathcal{G}^{(0)}_{\{k_i\}} + \Big{|} {1-\alpha_c z \over 1- \alpha_c} \Big{|}^2 \,\widetilde{\mathcal{G}}_{\{k_i\}}  (z, \bar{z}; \alpha_c, \bar{\alpha}_c)\right] \, ,
\fe
where $k_i$'s are the conformal dimensions of the operators, and $k_{ij}^- = k_i - k_j$ and $k^+_{ij\ldots l} = k_i + k_j + \ldots +k_l$. The cross ratios are defined as
\ie
|\xi_{ij}|^2 = {|z_{ij}|^2 \over t_{ij}^2}\, , \qquad \alpha_c = {A_1 \cdot A_3 A_2 \cdot A_4  \over A_1 \cdot A_4 A_2 \cdot A_3} \, , \qquad z = {z_{14} z_{23} \over z_{13} z_{24} }\, ,
\fe
and $z_{ij} = z_i - z_j$ and $t_{ij} = t_i -t_j$. To describe the R-symmetry group $SU(2)_L \times SU(2)_R$ of the CFT (or equivalently the isometry of $S^3$), we have associated each operator with spinors $A_i^{\alpha}, \bar{A}_i^{\dot \alpha}$, or equivalently a $SO(4)$ null vector $t_{i}^{\mu} = (\sigma_{\alpha \dot \alpha})^{\mu} A_i^{\alpha} \bar{A}_i^{\dot \alpha}$.  The role of these R-symmetry factors will become more explicit when we consider concrete examples.

The separation of the full correlator into $\mathcal{G}^{(0)}_{\{k_i\}}$ and $\widetilde{\mathcal{G}}_{\{k_i\}}$ such that  $\mathcal{G}^{(0)}_{\{k_i\}}$ is a simple rational function of cross ratios, whereas $\widetilde{\mathcal{G}}_{\{k_i\}}$ is the ``dynamical part", which contains the so-called $D$-functions and contributes non-trivially to Mellin amplitudes, which will be our main focus. Finally, the prefactor $\Big{|} {1-\alpha_c z \over 1- \alpha_c} \Big{|}^2$ is fixed by the half maximal supersymmetry and super conformal symmetry of the theory \cite{Giusto:2018ovt, Rastelli:2019gtj}.  We will call $\widetilde{\mathcal{G}}_{\{k_i\}}$ as the reduced correlator, whereas the full result including the prefactor $\Big{|} {1-\alpha_c z \over 1- \alpha_c} \Big{|}^2$ as the non-reduced correlator \footnote{The complete non-reduced correlator should also include the rational function part, $\mathcal{G}^{(0)}_{\{k_i\}}$. We will be mostly concerned with the non-rational terms and their corresponding Mellin amplitudes.}. 

To study holographic correlators, it is convenient to express them in Mellin space when it is possible. The Mellin amplitude of the connected part of a correlation function is defined through a Mellin transform \cite{Mack:2009mi, Penedones:2010ue}, 
\begin{align} \label{eq:Mellin}
\widetilde {\mathcal{G}}_{ \{k_i \}}(U,V)=&%\, {\pi  \over 2}  
\int  {ds \over 4\pi i}{dt \over 4\pi i} \,U^{{s - k^+_{34} \over 2} +L} V^{{t - {\rm min} \{ k^+_{23},\, k^+_{14} \} \over 2} }   \widetilde{\Gamma}_{\{k_i\}} (s, t)\, \widetilde{\mathcal{M}}_{ \{k_i \}} (s,t)\,\, ,
\end{align} 
where $\widetilde{\mathcal{M}}_{ \{k_i \}}(s, t)$ is the reduced Mellin amplitude, $\tilde{u}= \sum_{i=1}^4  k_i - s - t -2$,  the cross ratios $U, V$ are defined as
\ie
U&= (1-z) (1-\bar{z})\, , \qquad \quad V = z \bar{z} \, .
\fe and $ \widetilde{\Gamma}_{\{k_i\}} (s, t)$ is a product of $\Gamma$-functions
\ie \label{eq:wgamma}
 \widetilde{\Gamma}_{\{k_i\}} (s, t) = \Gamma\left({k^+_{12}-s \over 2}\right)\Gamma\left({ k^+_{34}-s \over 2}\right)   \Gamma\left({ k^+_{14}-t \over 2}\right) \Gamma\left({ k^+_{23}-t \over 2}\right) \Gamma\left({ k^+_{13}-\tilde{u} \over 2}\right)\Gamma\left({ k^+_{24}-\tilde{u} \over 2}\right) \, .
\fe
Finally, $L= k_4$ if $k^+_{14} \leq k^+_{23}$, and $L={k^+_{234} - k_1\over 2}$ if $k^+_{14} > k^+_{23}$. Without losing generality, we will only focus on the first case, namely $k^+_{14} \leq k^+_{23}$. 

One may put back the factor $\Big{|} {1-\alpha_c z \over 1- \alpha_c} \Big{|}^2$ in \eqref{eq:def} to obtain the non-reduced correlator. To study the Mellin amplitude, we express his factor in terms of cross ratios $U$ and $V$,
\ie \label{eq:R-factor}
\Big{|} {1-\alpha_c z \over 1- \alpha_c} \Big{|}^2 = \frac{1}{2} (\tau-\sigma+1)+\frac{U}{2} (\sigma +\tau
   -1)+\frac{V}{2}(\sigma -\tau
   +1)  - \frac{ \tau}{2}  (\alpha_c - \bar{\alpha}_c)(z- \bar{z})\, ,
\fe
where 
\ie \label{eq:stau}
\sigma&= {\alpha_c\, \bar{\alpha}_c \over (1- \alpha_c)(1 - \bar{\alpha}_c)} \, , \qquad \tau = {1\over (1- \alpha_c)(1 - \bar{\alpha}_c) } \, . 
\fe
The factor $(\alpha_c - \bar{\alpha}_c)(z- \bar{z})$ (which we will encounter again later) contains square roots in terms of cross ratios $U, V$, and it is incompatible with the definition of Mellin amplitudes we used in \eqref{eq:Mellin}. So we will have to drop this piece when we consider the Mellin amplitude for non-reduced correlator.\footnote{This can be achieved by summing over the correlator and its conjugate, namely $\alpha_c \leftrightarrow \bar{\alpha}_c$.} The same consideration was also used in \cite{Rastelli:2019gtj}. In general, the non-reduced Mellin amplitude ${\mathcal{M}}_{ \{k_i \}} (s,t)$ is defined as
\begin{align} \label{eq:Mellin2}
{\mathcal{G}}_{ \{k_i \}}(U,V)=&%\, {\pi  \over 2}  
\int  {ds \over 4\pi i}{dt \over 4\pi i} \,U^{{s - k^+_{34} \over 2} +L} V^{{t - {\rm min} \{ k^+_{23},\, k^+_{14} \} \over 2} }  \Gamma_{\{k_i \}}(s,t) \,   {\mathcal{M}}_{ \{k_i \}} (s,t)\,\, ,
\end{align} 
with $u= \sum_{i} k_i - s- t$ and
\ie  \label{eq:gamma}
 \Gamma_{\{k_i \}}(s,t)  = \Gamma\left({k^+_{12}-s \over 2}\right)\Gamma\left({ k^+_{34}-s \over 2}\right) \Gamma\left({ k^+_{14}-t \over 2}\right) \Gamma\left({ k^+_{23}-t \over 2}\right) \Gamma\left({ k^+_{13}- {u} \over 2}\right)\Gamma\left({ k^+_{24}- {u} \over 2}\right)\, .
\fe

From the definition of Mellin amplitudes in \eqref{eq:Mellin} and \eqref{eq:Mellin2}, it is straightforward to see the translation between the reduced Mellin amplitude and the non-reduced Mellin amplitude takes the following form,   
\ie \label{eq:MtM}
U^m V^n \widetilde{\mathcal{M}}_{ \{k_i\}  } (s,t) \rightarrow  {\widetilde{\Gamma}_{\{k_i\}}(s-2m,t-2n) \over  \Gamma_{\{k_i \}}(s,t ) }  \widetilde{\mathcal{M}}_{ \{k_i\} }  (s-2m, t-2n) \, .
\fe
Use the result of the above relation and \eqref{eq:R-factor} (after dropping the term $(\alpha_c - \bar{\alpha}_c)(z- \bar{z})$), we see that, for holographic correlators in $AdS_3 \times S^3$, the non-reduced Mellin amplitude $\mathcal{M}_{ \{k_i\} }  (s, t)$ is related to the reduced amplitude $\widetilde{\mathcal{M}}_{ \{k_i\} }(s, t)$ through the following relation, 
\ie  \label{eq:non-red2}
\mathcal{M}_{ \{k_i\} }  (s, t) &=\,  \frac{1}{8} (\tau -\sigma +1)(k_{13}^+-u) ( k_{24}^+-u) \, \widetilde{\mathcal{M}}_{ \{k_i\}  } (s,t) \cr
   &+ \, \frac{1}{8} (\sigma +\tau -1) (k_{12}^+ -s) (k_{34}^+ -s)\,  \widetilde{\mathcal{M}}_{ \{k_i\}  } (s{-}2,t) \cr
   &+\, \frac{1}{8} (\sigma -\tau +1)(k_{14}^+-t) (k_{23}^+-t) \,  \widetilde{\mathcal{M}}_{ \{k_i\} } (s,t{-}2) \, .
\fe

Let us now be concrete by considering the holographic correlators of CPOs in tensor multiplet. In tensor multiplet, there is a family of CPOs, $s_k$, with holomorphic and antihomorphic conformal dimensions $(h, \bar{h})=(k/2, k/2)$ with $k=1,2, \cdots$, so that it has dimension $h+\bar{h}=k$. They are in the $(j, \bar{j})=(h, \bar{h})$ representation of R-symmetry group $SU(2)_L \times SU(2)_R$. The operator $s_k$, after encoding the R-symmetry group factor, is then defined as
\ie
s_k(z_i, \bar{z}_i, t_i) = t_{i \mu_1} \ldots t_{i \mu_k} s_k^{\mu_1 \ldots \mu_k} (z_i, \bar{z}_i)\, ,
\fe
where $t_{i \mu}$ is the $SO(4)$ null vector associated with R-symmetry of the theory, as we introduced earlier.  The hidden conformal symmetry that we discussed earlier implies that the four-point correlator of $\langle s_{k_1} s_{k_2} s_{k_3} s_{k_4} \rangle$ is described by a 6D correlator of scalar operators \cite{Rastelli:2019gtj, Giusto:2020neo}
\begin{align} \label{eq:tensor}
\widetilde{\mathcal{G}}_{ s_{k_1} s_{k_2} s_{k_3} s_{k_4} } = {t_{12}^2 t_{34}^2} |z_{13}^2| |z_{24}^2| 
{|\xi_{12}|^{k_{12}^+ + k^-_{43}} |\xi_{34}|^{2k_4} \over |\xi_{13}|^{k^-_{21}+k^-_{43}} |\xi_{23}|^{k^-_{43} -k^-_{21}}} {g(Z_i) \over |Z_{12}|^4 |Z_{34}|^4} {\bigg \vert}_{t_1^{k_1} t_2^{k_2} t_3^{k_3} t_4^{k_4}}\, .
\end{align}
%where $f(Z)$ is determined by the correlator of lowest-weight operators with $k_i=1$, which will be given later. 
Importantly, $Z_i$ is a 6D coordinate by combining the $AdS_3$ and $S^3$ coordinates, 
\ie
Z_i = (z_i , \bar{z}_i, t_i^{\mu})\, ,
\fe
therefore $|Z_{ij}|^2 = |Z_i - Z_j|^2= |z_{ij}|^2  + t_{ij}^2$.  To obtain the correlator $\langle s_{k_1} s_{k_2} s_{k_3} s_{k_4}\rangle$, we simply Taylor expand $\widetilde{\mathcal{G}}_{ s_{k_1} s_{k_2} s_{k_3} s_{k_4} }$ and collect all the terms of the order $t_1^{k_1} t_2^{k_2} t_3^{k_3} t_4^{k_4}$, as indicated in the subscript in \eqref{eq:tensor}. Finally, $g(Z_i)$ is a function of cross ratios built out of 6D coordinates $Z_i$, which is the consequence of the hidden conformal symmetry. 

\subsection{Solution to the recursion relation}

To extract the coefficient of $t_1^{k_1} t_2^{k_2} t_3^{k_3} t_4^{k_4}$ efficiently, we rescale $t_{ij}^2 \rightarrow a_i a_j t_{ij}^2$, and express the $\widetilde{\mathcal{G}}_{ s_{k_1} s_{k_2} s_{k_3} s_{k_4} } $ as a contour integral following the ideas of \cite{Caron-Huot:2018kta}, 
\begin{align}
\widetilde{\mathcal{G}}_{ s_{k_1} s_{k_2} s_{k_3} s_{k_4} }  =\oint_{a_i=0} da_i a_i^{-k_i} {t_{12}^2 t_{34}^2} |z_{13}^2| |z_{24}^2| 
{|\xi_{12}|^{k_{12}^+ + k^-_{43}} |\xi_{34}|^{2k_4} \over |\xi_{13}|^{k^-_{21}+k^-_{43}} |\xi_{23}|^{k^-_{43} -k^-_{21}}} {g(Z_i(a_i)) \over |z_{12}^2 + a_1 a_2 t_{12}^2|^2 |z^2_{34} + a_3 a_4 t_{34}^2|^2}\, .
\end{align}
After a suitable rescaling on the integration variables, we can further simplify the result and express it in terms of cross ratios, \footnote{With a bit abuse of terminology, here we refer the relation \eqref{eq:recten} (as well as its generalisation when the gravity multiplet is included, as we will discuss in the next section) as a recursion relation, although it is not a relation that relates the correlator $\widetilde{\mathcal{G}}_{ s_{k_1} s_{k_2} s_{k_3} s_{k_4} }$ with the correlator of operators with neighboring weights. }

\begin{align} \label{eq:recten}
\widetilde{\mathcal{G}}_{ s_{k_1} s_{k_2} s_{k_3} s_{k_4} }   = \sigma ^{k_2-1} \tau ^{\frac{k^-_{12}- k^-_{34}}{2} }
   U^{\frac{k_{12}^+ +k^-_{43}-4}{2} } V^{\frac{k^-_{21}+k^-_{34}}{2}}   \oint_{a_i=0} da_i a_i^{-k_i}   
 {g\left( U' ,  V' \right) \over (1 + {a_1 a_2 \over U\, \sigma }  )^2 (1 + a_3 a_4 )^2 }  \, ,
 \end{align} 
where we have introduced the rescaled cross ratios $U', V'$, which are defined as
\begin{align} \label{Uprime}
U' = U { (1 + {1 \over \sigma   U} a_1 a_2 )(1 + a_3 a_4) \over (1 +a_1 a_3)(1+ a_2 a_4) } \,  ,  \qquad
V' = V { (1 + {\tau \over \sigma V} a_2 a_3 )(1 + a_1 a_4) \over (1 + a_1 a_3)(1+ a_2 a_4) }  \,  .
\end{align}
The function $g\left( U ,  V \right)$ that serves as initial data of the recursion relation is determined by the correlator of lowest-weight operators with $k_i=1$, and is given by \footnote{This correlator was first obtained in \cite{Giusto:2018ovt} by taking a limit on the heavy-heavy-light-light correlators computed in \cite{Galliani:2017jlg, Bombini:2017sge}. }
\begin{align}
g\left( U ,  V \right) = U\! \int {ds d t \over (4\pi i)^2} U^{s \over 2} V^{{t\over 2} -1}  
\Gamma^2(1- {s \over 2} ) \Gamma^2(1- {t \over 2} ) \Gamma^2(1- {\tilde{u} \over 2 }) \left( \frac{\delta_{f_1 f_2} \delta_{f_3 f_4} }{s} + \frac{\delta_{f_2 f_3} \delta_{f_1 f_4}}{t} + \frac{\delta_{f_1 f_3} \delta_{f_2 f_4}}{\tilde{u}}   \right)\, ,
\end{align}
with $\tilde{u} = 2-s-t$. Here we have expressed the function in Mellin space, which can be straightforwardly re-expressed in the coordinate space in terms of $D$-functions \cite{Giusto:2018ovt}.

Perform the contour integral using $(1+x)^n = \sum_{i=0}^{\infty} {\Gamma(i-n) \over \Gamma(-n)} {(-x)^i \over i!}$, and shift the integration variables by $s \rightarrow s + 2+2m_1 -k_{12}^+$ and $t \rightarrow t + 2+ 2 m_5 - k_{23}^+$, we arrive at
\ie
\widetilde{\mathcal{G}}_{ s_{k_1} s_{k_2} s_{k_3} s_{k_4} } &= \int  {ds dt \over (4\pi i)^2} 
\sum^{\infty}_{m_1=0, m_2=0}\sigma ^{m_2 + {k^-_{21} + k^-_{43} \over 2}} \tau ^{k_1-1 - m_{12}}
   U^{{s + k^-_{43}\over 2}  } V^{{t-k_{14}^+\over 2} } \, \widetilde{\Gamma}_{\{k_i \}}(s,t)  \\
   &  \times     
     {1 \over \prod_{i=1}^6 m_i!}  
  \left( \frac{\delta_{f_1 f_2} \delta_{f_3 f_4} }{s + 2+2m_1 -k_{12}^+ } + \frac{\delta_{f_2 f_3} \delta_{f_1 f_4}}{t + 2+ 2 m_5 - k_{23}^+ } + \frac{\delta_{f_1 f_3} \delta_{f_2 f_4}}{ \tilde{u} +2+ 2m_2-k_{13}^+ }   \right)  \, ,  
\fe
where $\widetilde{\Gamma}_{\{k_i \}}(s,t)$ is given in \eqref{eq:wgamma}, and
$m_i$ for $i>2$ are determined in terms of $m_1, m_2$, 
\ie
m_3 &=  k_1 - m_{12} -1 \, , \qquad m_4 =  {k^-_{31} + k^-_{42} \over 2}   + m_1 \, , \cr
m_5 &=  {k_{12}^+ + k^-_{34}  \over 2}  - m_{12} -1 \, , \qquad m_6 ={k^-_{21} +k^-_{43} \over 2} +m_2 \, ,
  \fe
with $m_{ij}=m_i +m_j$. Note, the summation on $m_1, m_2$ is truncated due to the factorials $m_i!$ in the denominator. 
According to the definition of Mellin amplitudes given in \eqref{eq:Mellin}, we conclude that the reduced Mellin amplitude of $\langle s_{k_1} s_{k_2} s_{k_3} s_{k_4}\rangle$ is given by
\ie \label{eq:non-red}
\widetilde{\mathcal{M}}_{ s_{k_1} s_{k_2} s_{k_3} s_{k_4} } (s, t)&= \sum^{\infty}_{m_1=0, m_2=0}  
{\sigma ^{m_2 + {k^-_{21} + k^-_{43} \over 2}} \tau ^{k_1-1 - m_{12}}  \over \prod_{i=1}^6 m_i!}    \cr
& \times \left( \frac{\delta_{f_1 f_2} \delta_{f_3 f_4}}{s + 2+2m_1 -k_{12}^+ } + \frac{\delta_{f_2 f_3} \delta_{f_1 f_4}}{t + 2+ 2 m_5 - k_{23}^+ } + \frac{\delta_{f_1 f_3} \delta_{f_2 f_4}}{ \tilde{u} +2+ 2m_2-k_{13}^+ }   \right) \, . 
\fe
For the special case $k_1=k_2=k$ and $k_3=k_4=\ell$, we find the formula is in agreement with the result given in \cite{Giusto:2019pxc} (up to an overall factor of $-2k\ell$ due to a different convention we use here).

We conclude this section by taking two interesting limits of the Mellin amplitudes: flat-space limit and maximally R-symmetry violating (MRV) limit. We will consider the limits on the full non-reduced Mellin amplitudes using \eqref{eq:non-red2}. Flat-space limit is achieved by setting $s \rightarrow \infty, t \rightarrow \infty$, which yields
\ie \label{eq:flat0}
\mathcal{M}_{s_{k_1} s_{k_2} s_{k_3} s_{k_4} }  (s,t)  \rightarrow  P_{\{k_i\}} (\sigma, \tau)  \left( u\,t+s\,t\,\s+s\,u\,\tau \right)  \left( \frac{\delta_{f_1 f_2} \delta_{f_3 f_4} }{s} + \frac{ \delta_{f_1 f_4} \delta_{f_2 f_3} }{t} + \frac{\delta_{f_1 f_3} \delta_{f_2 f_4} }{u }  \right) \, ,
\fe
where $u \rightarrow -s-t$ in the flat-space limit, and the overall factor $P_{\{k_i\}} (\sigma, \tau)$ is given by
\begin{align}
P_{\{k_i\}}(\sigma, \tau)&=- {1\over 4} \sum^{\infty}_{m_1=0, m_2=0}  
{\sigma ^{m_2 + {k^-_{21} + k^-_{43} \over 2}} \tau ^{k_1-1 - m_{12}}  \over \prod_{i=1}^6 m_i!} \, .
\end{align}
One may perform one of the summations in $P_{\{k_i\}}(\sigma, \tau)$ and express the result in terms of a Hypergeometric function. We see that \eqref{eq:flat0} is in agreement with the flat-space amplitude given in \eqref{eq:flat}. The factor $(u\,t+s\,t\,\s+s\,u\,\tau)$, arising from $\Big{|} {1-\alpha_c z \over 1- \alpha_c} \Big{|}^2$ as can be seen from \eqref{eq:non-red2}, represents the fact that the theory has half maximal supersymmetry factor. In the case of maximal supersymmetric theories, it is $(u\,t+s\,t\,\s+s\,u\,\tau)^2$ that associates with the supersymmetry \cite{Chester:2018dga, Alday:2020lbp}.  %&= \sigma ^{\frac{1}{2} (k_{21}^{-}-{k_{34}^-})} \sum_{m_1=0}^{\infty} \tau ^{-{m_1}+{k_1}-1} \nonumber\\
%&\times \frac{\, _2\tilde{F}_1\left({m_1}-{k_1}+1,\frac{1}{2} (2 {m_1}-k_{12}^{+}-k_{34}^{-}+2);\frac{1}{2} (k_{21}^{-}-k_{34}^{-}+2);\frac{\sigma }{\tau }\right)}{\Gamma ({m_1}+1) \Gamma ({k_1}-{m_1}) \Gamma \left(\frac{1}{2} (-2 {m_1}+k_{12}^{+}+k_{34}^{-})\right) \Gamma \left(\frac{1}{2} (2 {m_1}-k_{12}^{+}+k_{34}^{+}+2)\right)}\, ,
%\end{align}
%where $_2\tilde{F}_1$ is the regularised Hypergeometric function. 

We now consider the MRV limit. This was first introduced in \cite{Alday:2020lbp, Alday:2020dtb} for the study of holographic correlators in maximal supersymmetric theories. The limit chooses to align the R-symmetry directions $A^{\alpha}_i$ such that the $u$-channel contribution vanishes, namely we set $A_1 \cdot A_3=A_2 \cdot A_4=0$ and $\bar{A}_1 \cdot \bar{A}_3=\bar{A}_2 \cdot \bar{A}_4=0$, which implies $\alpha_c = \bar{\alpha}_c=0$. In terms of $\s$ and $\tau$,  we have $\s \rightarrow 0,\, \tau \rightarrow 1$ in the MRV limit. So in this limit, the factor $(\alpha_c - \bar{\alpha}_c)(z-\bar{z})$ in \eqref{eq:R-factor} drops out, and the non-reduced Mellin amplitude is always well-defined. 

For the correlator given in \eqref{eq:non-red}, we see that in the MRV limit only $m_2= -\frac{k^-_{21} + k^-_{43}}{2}$ term of the $m_2$ sum in (\ref{eq:non-red})  contributes, therefore we have
\begin{align} 
\mathcal{M}_{{\rm MRV},\, s_{k_1} s_{k_2} s_{k_3} s_{k_4} } (s, t)&=\frac{1}{4}\, (s+t-k_{13}^+)(s+t-k_{24}^+)   \\
&\times \sum_{m_1=0}^{\infty} \left( \frac{\delta_{f_1 f_2} \delta_{f_3 f_4}}{s+2+2m_1-k_{12}^+}+\frac{\delta_{f_2 f_3} \delta_{f_1 f_4}}{t-2m_1+k_{23}^-}+\frac{\delta_{f_1 f_3} \delta_{f_2 f_4}}{-s-t+k_{13}^+} \right) \nonumber \\
&\times\frac{1}{\Gamma(m_1+1)\Gamma(k_2-m_1) \Gamma(\frac{k_{12}^- + k_{34}^-}{2}+1)\Gamma(\frac{k_{12}^+ - k_{34}^-}{2}-m_1)\Gamma(m_1+\frac{k_{34}^+-  k_{12}^+}{2} +1 )} \, . \nonumber
\end{align}
We note that the $u$-channel poles are cancelled by the prefactor $(s+t-k_{13}^+)$, and there is a zero in $u$-channel when $s+t-k_{24}^+=0$. These are the properties of holographic correlators in the MRV limit that played key roles for constructing tree-level holographic correlators in other $AdS$ backgrounds with maximal supersymmetry, which include $AdS_5 \times S^5, AdS_4 \times S^7$, and $AdS_7 \times S^4$ \cite{Alday:2020lbp, Alday:2020dtb}.  

So far we have considered the MRV limit with the $u$-channel R-symmetry spinors $A_{1}, A_{3}$ (and $A_{2}, A_{4}$) being aligned.  One may also consider the MRV limit with the $s$-channel R-symmetry spinors $A_{1}, A_{2}$ (and $A_{3}, A_{4}$) as well as their conjugates being aligned. With this choice, we have $\alpha_c=\bar{\alpha}_c=1$, and according to \eqref{eq:stau}, $\sigma = \tau \rightarrow \infty$ . Therefore, in this limit, the terms with the highest degree in $\sigma, \tau$ dominate, and we find the Mellin amplitude takes the following form, 
\ie
\mathcal{M}_{{\rm MRV'},\, s_{k_1} s_{k_2} s_{k_3} s_{k_4} } (s, t)&=\frac{1}{4}\,\s^{\frac{k_{12}^+-k_{43}^-}{2}}\, (s- k_{12}^+) (s-k_{34}^+)   \\
&\times \sum_{m_2=0}^{\infty} \left( \frac{\delta_{f_1 f_2} \delta_{f_3 f_4}}{s-k_{12}^+}+\frac{\delta_{f_2 f_3} \delta_{f_1 f_4}}{t+k_{23}^-}+\frac{\delta_{f_1 f_3} \delta_{f_2 f_4}}{-s-t+k_{13}^+} \right)   \\
&\times\frac{1}{\Gamma(m_2+1) \Gamma ({k_1}-{m_2}) \Gamma(\frac{k_{34}^+-k_{12}^+}{2} +1) \Gamma({m_2}+\frac{k_{24}^+-k_{13}^+}{2}) \Gamma (-m_2+\frac{k_{12}^+-k_{34}^-}{2})} \, . 
\fe
We see that in this choice of the MRV limit, there is no $s$-channel pole and has a zero at $(s-k_{34}^+) =0$, as expected.

%%%%%
\section{Four-point correlators of operators in tensor and gravity multiplets}
\label{secgravity}
%%%%%

\subsection{Flat space superamplitudes}

We will now consider the correlators involving operators in  gravity multiplet. As we commented, compared to the correlators of operators in tensor multiplet, these correlators are more involved. Let us begin with the amplitudes in flat-space. The four-point superamplitude in 6D $(2,0)$ supergravity of general external states is given by \cite{Heydeman:2018dje}
\begin{equation}
  \label{eq:fs4p}
  \mathcal{A}_4 = G_6 \delta^8(\sum_{i=1}^4 q_i) \delta^6( \sum_{i=1}^4 p_i) \frac{[1_{\hat a_1} 2_{\hat a_2} 3_{\hat a_3} 4_{\hat a_4}][1_{\hat b_1} 2_{\hat b_2} 3_{\hat b_3} 4_{\hat b_4}] }{\bold{s} \,  \bold{t}\, \bold{u}}\, .
\end{equation}
To simplify the discussion, we have assumed that the tensors have the same flavour, and more general cases can be found in \cite{Heydeman:2018dje}, which are constructed using twistor formulation.  The square parenthesis  is defined as $[i_{\hat{a}_1} j_{\hat a_2} k_{\hat a_3} l_{\hat a_4}]:= \epsilon^{ABCD} \tilde{\lambda}_{i \, A\, \hat{a}_1}  \tilde{\lambda}_{j \, B\, \hat{a}_2}  \tilde{\lambda}_{k \, C\, \hat{a}_3} \tilde{\lambda}_{l \, D\, \hat{a}_4}$. Importantly, as pointed out in \cite{Giusto:2020neo}, after stripping of the delta-function prefactors $G_6 \delta^8(\sum_{i=1}^4 q_i) \delta^6( \sum_{i=1}^4 p_i)$, this general four-point superamplitude in 6D $(2,0)$ supergravity is again invariant under 6D conformal transformation, which hints on hidden conformal symmetry for all four-point tree-level holographic correlators (instead of just those in tensor multiplet). 

The superamplitude of given states can be obtained from \eqref{eq:fs4p} by appropriately choosing the little group indices. For the states in tensor multiplet (they are the states without free little group indices), we contract all the little group indices, the numerator simplifies: 
\ie
{[1_{\hat a_1} 2_{\hat a_2} 3_{\hat a_3} 4_{\hat a_4} ]} [1^{\hat a_1} 2^{\hat a_2} 3^{\hat a_3} 4^{\hat a_4}]  = \bold{s}^2+\bold{t}^2+\bold{u}^2 \, ,
\fe 
we then find \eqref{eq:fs4p} reproduces \eqref{eq:flat} when the tensors have the same flavour, which is the case we consider here. We are interested in the amplitudes with two states in tensor multiplet and two in  gravity multiplet (the states with free little group indices), for which we have
\ie \label{eq:gravity}
  \mathcal{A}_4 = G_6 \delta^8(\sum_{i=1}^4 q_i) \delta^6( \sum_{i=1}^4 p_i)\frac{[1_{\hat a_1} 2_{\hat a_2} 3_{\hat a_3} 4_{\hat a_4}][1^{\hat a_1} 2^{\hat a_2} 3_{\hat b_3} 4_{\hat b_4}] }{ \bold{s}\, \bold{t} \, \bold{u}}\, ,
\fe
where we contact the little-group indices for the first two states (they are in tensor multiplet) and leave the indices free for the last two states (they are in gravity multiplet). Here we have also assumed two tensors have the same flavour, otherwise the amplitude vanishes, said in another way, we have suppressed a flavour factor $\delta_{f_1 f_2}$ in \eqref{eq:gravity}, with $f_1, f_2$ being the flavours of the tensors. 

For the comparison of holographic correlators in $AdS_3$, we compactify the above amplitude to three dimensions. This is done by reducing the 6D spinors to 3D spinor, which effectively sets $[1_{-} 2_{-} 3_+ 4_+] = - \langle 12\rangle \langle34 \rangle$, where in 3D, the massless momentum can be expressed as $p_{i}^{\alpha \beta} = \lambda_i^{\alpha} \lambda_i^{\beta}$, and the angle braket is defined as $ \langle ij\rangle = \lambda_i^{\alpha} \lambda_j^{\beta} \epsilon_{\alpha \beta}$, which relates to Mandelstam variables by $\langle ij\rangle^2=(p_i+p_j)^2$.  We then obtain the dimension reduced amplitude of four three-dimensional scalars from \eqref{eq:gravity}, given as
\ie \label{eq:flat1}
  \mathcal{A}_4 = G_6 \delta^8(\sum_{i=1}^4 q_i) \delta^6( \sum_{i=1}^4 p_i) \frac{ \bold{s}^2 -  \bold{t}^2 - \bold{u}^2}{\bold{s}\, \bold{t} \, \bold{u}}=  G_6 \delta^8(\sum_{i=1}^4 q_i) \delta^6( \sum_{i=1}^4 p_i) \frac{2}{ \bold{s}}\, .
\fe
The structure of this result is expected. When compactified to 4D, the $(2,0)$ supergravity becomes supersymmetric multiple-$U(1)$ Einstein-Maxwell theory. Both the scalar arising from the 6D graviton multiplet and the scalar from the 6D tensor multiplet are matters of Einstein-Maxwell theory \cite{Heydeman:2018dje}, but they belong to different $U(1)$'s of Maxwell theory, therefore they do not couple to a 4D graviton, which reflects in \eqref{eq:flat1} by the fact that there are no $t$- or $u$-channel poles. This fact implies only the $s$-channel pole is allowed, and a further reduction to 3D does not change the structure. 

\subsection{Correlators with gravity multiplet operators and hidden conformal symmetry} 

A family of CPOs in gravity multiplet that we will consider here are scalar operators which have left-right symmetry $(h, \bar{h}) = (k/2, k/2)$, with $k=2,3, \ldots$, we will denote them as $\sigma_k$. As we see, they have similar structures as the operators $s_k$ that we studied in the previous section. However, these operators arise from the Kaluza-Klein reduction of the supergravitons in 6D $(2,0)$ supergravity over the $S^3$,\footnote{See the Tables 1, 2, 3 in \cite{Rastelli:2019gtj} for more details of the spectrum in $AdS_3 \times S^3$ supergravity.} therefore the interaction couplings involving $\sigma_k$ operators are rather different from those of operators $s_k$.  For instance, a coupling of three $\sigma$'s is allowed but not for three $s$'s \cite{Arutyunov:2000by}.  As already indicated in the flat-space amplitudes \eqref{eq:fs4p}, these properties make the correlators in gravity multiplet much more complicated comparing to those in tensor multiplet. As in the case of $s_k$, to incorporate the R-symmetry we introduce the $SO(4)$ null vector  $t_{i\,\mu}$ in the definition of the operators, which are given by
\ie
 \sigma_k(z_i,\bar{z}_i;t_i) = t_{i\,\mu_1} \ldots t_{i\,\mu_{k}} \sigma_k^{\mu_1 \ldots \mu_{k}}(z_i,\bar{z}_i) \, .
\fe
As we will see that, unlike the correlators of $s_k$ that we studied in the previous section, even for the reduced correlators, when the operators $\sigma_k$ are involved, the correlators cannot be expressed in terms of $\sigma, \tau$ and $U, V$ only. To describe these correlators, it is necessary to use $\alpha_c, \bar{\alpha}_c$ and $z, \bar{z}$.  This should be closely related to the fact that the flat-space superamplitude involving graviton states, as given in \eqref{eq:gravity}, cannot be expressed in terms of Mandelstam variables only.

As we already anticipated earlier when we discussed the flat-space amplitudes, and as understood in \cite{Giusto:2020neo}, the holographic correlators $\langle s_{k_1} s_{k_2} \sigma_{k_3} \sigma_{k_4}\rangle$ also exhibit a hidden conformal symmetry, just as for the correlators in tensor multiplet. In particular, the holographic correlators for arbitrary conformal weights $k_i$ can be described by a single 6D CFT correlator with two scalar operators (corresponding to $s_{k_1}$ and $s_{k_2}$) and two $3$-forms (corresponding to $\sigma_{k_3}$ and $\sigma_{k_4}$). In practice, this leads to a recursion relation that determines $\langle s_{k_1} s_{k_2} \sigma_{k_3} \sigma_{k_4}\rangle$ for any $k_i$ in terms of initial datas which can be obtained from four-point correlators of operators with low conformal weights. Explicitly, the correlator  $\langle s_{k_1} s_{k_2} \sigma_{k_3} \sigma_{k_4}\rangle$ is described by \cite{Giusto:2020neo}
\begin{align} \label{eq:rec}
\widetilde{\mathcal{G}}_{s_{k_1} s_{k_2} \sigma_{k_3} \sigma_{k_4} } &={t_{12}^2 t_{34}^2} |z_{13}^2| |z_{24}^2| 
{|\xi_{12}|^{k_{12}^+ + k^-_{43}} |\xi_{34}|^{2k_4} \over |\xi_{13}|^{k^-_{21}+k^-_{43}} |\xi_{23}|^{k^-_{43} -k^-_{21}}} 
\frac{1}{|Z_{12}|^4 |Z_{34}|^8}\Bigl\{g_1(Z)\, t_{34}^2 |z_{34}|^2\cr
&+ g_2(Z) \Bigl[  \left(t_{14}^2 t_{23}^2 - t_{13}^2 t_{24}^2 \right) \frac{|z_{34}|^2}{|Z_{12}|^2} + t_{34}^2 \frac{|z_{14}|^2 |z_{23}|^2-|z_{13}|^2 |z_{24}|^2}{|Z_{12}|^2}\Bigr]\cr
 &+ g_3(Z) \Bigl[ \frac{(t_{13}^2 t_{24}^2+t_{14}^2t_{23}^2)|z_{34}|^2+t_{34}^2 (|z_{14}|^2 |z_{23}|^2+|z_{13}|^2|z_{24}|^2)}{2 |Z_{12}|^2}\\
  & - \frac{t_{12}^2 t_{34}^2 |z_{13}|^2 |z_{24}|^2 (z+\bar z)}{|Z_{12}|^4} - \frac{t_{13}^2 t_{14}^2 |z_{23}|^2 |z_{24}|^2 +t_{23}^2 t_{24}^2 |z_{13}|^2 |z_{14}|^2 }{|Z_{12}|^4} \cr
 & - \frac{t_{13}^2 t_{24}^2 (|z_{12}|^2 |z_{34}|^2- |z_{13}|^2 |z_{24}|^2)+t_{14}^2 t_{23}^2 (|z_{12}|^2 |z_{34}|^2- |z_{14}|^2 |z_{23}|^2)}{|Z_{12}|^4} \cr
 & - 4\, \epsilon_{\mu_1 \mu_2 \mu_3 \mu_4} t_1^{\mu_1} t_2^{\mu_2} t_3^{\mu_3} t_4^{\mu_4} (z-\bar z)  \frac{|z_{13}|^2 |z_{24}|^2}{|Z_{12}|^4} \Bigr] \Bigr\}{\bigg \vert}_{t_1^{k_1} t_2^{k_2} t_3^{k_3} t_4^{k_4}} \, ,  \nonumber
\end{align}  
% \newpage
%\!\!\!\!\!\!\!\!\!\!\! 
where, again, the correlator $\langle s_{k_1} s_{k_2} \sigma_{k_3} \sigma_{k_4}\rangle$ is obtained by Taylor expanding the above expression to the order $t_1^{k_1} t_2^{k_2} t_3^{k_3} t_4^{k_4}$. We have suppressed the flavour factor $\delta_{f_1 f_2}$ of the tensors in the above expression.  

Compare to the correlators of operators in tensor multiplet, we now have more unknown functions that is because, unlike the scalars, the self-dual $3$-forms in 6D allow for more independent structures. In particular, it was found there are three of them \cite{Giusto:2020neo}, which are associated with the unknown functions $g_1, g_2$ and $g_3$ in \eqref{eq:rec}. These functions that serve as initial data of the recursion relation are determined by comparing with the known results of correlators $\langle s_{1} s_{1} \sigma_{2} \sigma_{2}\rangle$ and $\langle s_{2} s_{2} \sigma_{2} \sigma_{2}\rangle$, and they are given by \cite{Giusto:2020neo}
\begin{align} \label{eq:initial}
g_1(U, V) &=  \int {ds dt \over (4\pi i)^2}  U^{s \over 2} V^{{t\over 2} -2} \frac{(s-6) (s-4)}{6 (s-2)}\Gamma^2 (2-\frac{s}{2}) \Gamma^2 (2-\frac{t}{2}) \Gamma^2 (\frac{s+t}{2}-2 ) \, , \\
g_2(U, V) &=  \int {ds dt \over (4\pi i)^2}  U^{s \over 2} V^{{t\over 2} -2}  \frac{(s-6) (s-4) (s+2 t-8)}{12 (t-2) (s+t-6)} \Gamma^2 (2-\frac{s}{2}) \Gamma^2 (2-\frac{t}{2}) \Gamma^2 (\frac{s+t}{2}-2 ) \, , \cr
g_3(U, V) &=  \int {ds dt \over (4\pi i)^2}  U^{s \over 2} V^{{t\over 2} -2}  \frac{(s-6) (s-4)^2}{6 (t-2) (s+t-6)} \Gamma^2 (2-\frac{s}{2}) \Gamma^2 (2-\frac{t}{2}) \Gamma^2 (\frac{s+t}{2}-2 )\, .  \nonumber
\end{align}

It is worth noting that not any initial data (namely $g_1, g_2, g_3$ given in \eqref{eq:initial}) would generate sensible correlation functions of operators with higher conformal weights. Therefore, we expect the recursion relation can even constrain the initial data. Surprisingly, we find that the initial data is in fact uniquely fixed by simple consistency conditions of correlators generated from the recursion relation. Concretely, we begin by assuming the following ansatzs for the initial data, 
\begin{align} \label{eq:ansatz}
g_1(U, V) &=  \int {ds dt \over (4\pi i)^2}  U^{s \over 2} V^{{t\over 2} -2} \frac{ \sum a_{i, j}s^i t^j }{s-2}\Gamma^2 (2-\frac{s}{2}) \Gamma^2 (2-\frac{t}{2}) \Gamma^2 (\frac{s+t}{2}-2 ) \, , \\
g_2(U, V) &=  \int {ds dt \over (4\pi i)^2}  U^{s \over 2} V^{{t\over 2} -2} \left(
 \frac{\sum  b_{i, j}s^i t^j }{t-2} +  \frac{\sum  c_{i, j}s^i t^j }{s+t-6} \right) \Gamma^2 (2-\frac{s}{2}) \Gamma^2 (2-\frac{t}{2}) \Gamma^2 (\frac{s+t}{2}-2 ) \, , \cr
g_3(U, V) &=  \int {ds dt \over (4\pi i)^2}  U^{s \over 2} V^{{t\over 2} -2}   \left(
 \frac{\sum  d_{i, j}s^i t^j }{t-2} +  \frac{\sum  e_{i, j}s^i t^j }{s+t-6} \right) \Gamma^2 (2-\frac{s}{2}) \Gamma^2 (2-\frac{t}{2}) \Gamma^2 (\frac{s+t}{2}-2 ) \, . \nonumber
\end{align}
Here the summations on $i, j$ are restricted by $i+j\leq 2$ due to the two-derivative power counting of supergravity. We have also used the fact that the functions should have correct pole structures, which are dictated by the exchanged states. The ansatzs contain $30$ free parameters, namely the coefficients $a_{ij}, b_{ij}, c_{ij}, d_{ij}$, and $e_{ij}$ in \eqref{eq:ansatz}.  We then require the full correlators of higher weights that are generated from the recursion relation by plugging the ansatzs \eqref{eq:ansatz} into \eqref{eq:rec} to have right pole structures and the correct power counting. In particular, we know that the correlator $M_{s_1 s_1  \sigma_k \sigma_k}(s, t)$ has and only has simple pole at $s=0$, since only the massless graviton and graviphoton are allowed to be exchanged in the $s$ channel.  For as a two-derivative theory, we also know that $M_{s_1 s_1  \sigma_k \sigma_k}(\beta s, \beta t) \sim \beta$ in the limit $\beta \rightarrow \infty$. By imposing such conditions for $k=2, 3$, we find that the anstazs of $g_1, g_2$ and $g_3$ given in \eqref{eq:ansatz} are uniquely fixed up to an overall factor, and agree precisely with what are given in \eqref{eq:initial} which were determined from explicit known results. 

\subsection{Solution to the recursion relation}

We will now solve the recursion relation following the same strategy of section \ref{sec:tensor-cor} for the simpler correlators of tensor multiplet. In particular, we express the recursion relation as contour integrals. We also note that, due to the last term in \eqref{eq:rec} proportional to $\epsilon_{\mu_1 \mu_2 \mu_3 \mu_4}$, it is not possible to express the correlator in terms of cross ratios $U, V$ and $\sigma, \tau$ only, instead it is necessary to use $z, \bar z$ and $\alpha_c, \bar \alpha_c$. Therefore it is natural to separate the correlator into two parts where one of them contains $z, \bar z$ and $\alpha_c, \bar \alpha_c$ (in the form of $(\alpha_c - \bar{\alpha}_c)(z-\bar{z})$, arising from the term that is proportional to $\epsilon_{\mu_1 \mu_2 \mu_3 \mu_4} t_1^{\mu_1}t_2^{\mu_2}t_3^{\mu_3}t_4^{\mu_4}$ in \eqref{eq:rec}), which we will denote as the chiral sector $\widetilde{\mathcal{G}}^{(c)}_{s_{k_1} s_{k_2}  \sigma_{k_3} \sigma_{k_4}}$,  and the remaining part only depends on $U, V$ and $\sigma, \tau$, which is the non-chiral sector $ \widetilde{\mathcal{G}}^{(nc)}_{s_{k_1} s_{k_2}  \sigma_{k_3} \sigma_{k_4}}$. Therefore, we will express the full correlator as
\begin{align} \label{eq:recurGR}
\widetilde{\mathcal{G}}_{s_{k_1} s_{k_2}  \sigma_{k_3} \sigma_{k_4}} &= \widetilde{\mathcal{G}}^{(c)}_{s_{k_1} s_{k_2} \sigma_{k_3} \sigma_{k_4} }  + \widetilde{\mathcal{G}}^{(nc)}_{s_{k_1} s_{k_2} \sigma_{k_3} \sigma_{k_4} }   \, , 
\end{align} 
where $\widetilde{\mathcal{G}}^{(c)}_{s_{k_1} s_{k_2} \sigma_{k_3} \sigma_{k_4} } $ and $\widetilde{\mathcal{G}}^{(nc)}_{s_{k_1} s_{k_2} \sigma_{k_3} \sigma_{k_4} } $ are given in terms of contour integrals, 
\begin{align}    \label{eq:Gcc}
\widetilde{\mathcal{G}}^{(c)}_{s_{k_1} s_{k_2}  \sigma_{k_3} \sigma_{k_4}}&= 
(\alpha_c -\bar \alpha_c)(z-\bar z) \sigma ^{k_2-2} \tau ^{\frac{k^-_{12}- k^-_{34}}{2} +1}
   U^{\frac{k_{12}^++k^-_{43}-8}{2} } V^{\frac{k^-_{21}+k^-_{34}}{2}}  \cr
& \times
 \oint_{a_i=0}  da_i\, a_1^{1-k_1} a_2^{1-k_2}   a_3^{1-k_3} a_4^{1-k_4} { g_3(U', V')   \over (1+ {a_1 a_2 \over U\, \sigma }  )^4 (1 + a_3 a_4 )^4 }  \, , 
 \end{align} 
 and
 \begin{align} \label{eq:Gnc}
\widetilde{\mathcal{G}}^{(nc)}_{s_{k_1} s_{k_2}  \sigma_{k_3} \sigma_{k_4}}&= 
\sigma ^{k_2-1} \tau ^{\frac{k^-_{12}- k^-_{34}}{2} }
   U^{\frac{k^+_{12}+k^-_{43}-4}{2} } V^{\frac{k^-_{21}+k^-_{34}}{2}}   \oint_{a_i=0}  da_i\, a_1^{-k_1} a_2^{-k_2}   a_3^{1-k_3} a_4^{1-k_4} 
{1  \over (1+ {a_1 a_2 \over U\, \sigma }  )^2 (1 + a_3 a_4 )^4 } \cr
\times &  \Bigl\{g_1(U', V')   + \frac{   a_1 a_2 \tau }{ (1 + {a_1 a_2 \over U\, \sigma }  )  U \sigma } F^{(s)}_2(U', V')  + \frac{   a_1 a_2  }{ (1 + {a_1 a_2 \over U\, \sigma }  )  U  } F^{(a)}_2(U', V') +
 \frac{   V }{ (1 + {a_1 a_2 \over U\, \sigma }  )  U  } F^{(s)}_2(U', V') 
 \cr
   +&\,
 \frac{ F^{(a)}_2(U', V')  }{ (1 + {a_1 a_2 \over U\, \sigma }  )  U  }  +   g_3(U', V') \Bigl[  -\frac{   a_1 a_2    (1-U+V)}{
  (1 + {a_1 a_2 \over U\, \sigma }  )^2 U^2 \sigma} -  \frac{a_1^2  v + a_2^2  \tau/\sigma }{ (1 + {a_1 a_2 \over U\, \sigma }  )^2 U^2}  -a_1 a_2   \frac{   ( U - 1)+ \tau/\sigma (U- V)}{ (1 + {a_1 a_2 \over U\, \sigma }  )^2  U^2 }  \Bigr] \Bigr\}  \, , 
\end{align} 
with $U', V'$ given in \eqref{Uprime}. Here we have expressed the results in the form of contour integrals, as we did in the previous section.  Finally, $F^{(s)}_2$ and $F^{(a)}_2$ are linear combinations of $g_2$ and $g_3$. In Mellin space, they are given by
\ie
F^{(s)}_2  &= g_3/2 +g_2=  \int {ds dt \over (4\pi i)^2}  U^{s \over 2} V^{{t\over 2} -2} \frac{(s-6) (s-4)}{6 (t-2)} \Gamma^2 (2-\frac{s}{2}) \Gamma^2 (2-\frac{t}{2}) \Gamma^2 (\frac{s+t}{2}-2 )  \, , \cr
 F^{(a)}_2 &= g_3/2 -g_2= - \int {ds dt \over (4\pi i)^2}  U^{s \over 2} V^{{t\over 2} -2} \frac{(s-6) (s-4)}{6 (s+t-6)} \Gamma^2 (2-\frac{s}{2}) \Gamma^2 (2-\frac{t}{2}) \Gamma^2 (\frac{s+t}{2}-2 ) \, .
\fe
In next subsections, we will solve the recursion relation for $\widetilde{\mathcal{G}}^{(c)}_{s_{k_1} s_{k_2}  \sigma_{k_3} \sigma_{k_4}}$ and $\widetilde{\mathcal{G}}^{(nc)}_{s_{k_1} s_{k_2}  \sigma_{k_3} \sigma_{k_4}}$, respectively. 

\subsubsection{The chiral sector} 
\label{sec:chiral}

The recursion relation for the chiral sector $\widetilde{\mathcal{G}}^{(c)}_{s_{k_1} s_{k_2}  \sigma_{k_3} \sigma_{k_4}}$ as given in \eqref{eq:Gcc} is relatively simple. It has an analogous structure as the correlators of tensor multiplet.  After factoring out $(\alpha_c -\bar \alpha_c)(z-\bar z)$, we can then formally express $\widetilde{\mathcal{G}}^{(c)}_{s_{k_1} s_{k_2}  \sigma_{k_3} \sigma_{k_4}}$  in Mellin space using the standard definition in \eqref{eq:Mellin}, and the recursion relation can be solved in a similar manner as for the correlators of tensor multiplet that we studied in the previous section. Since the computation is very similar to that has been done for  the correlators of tensor multiplet, we will not repeat the steps here. Instead we will simply present the final result of this particular part of the correlator, which is given by
          \begin{align} \label{eq:chiral}
{\widetilde{\mathcal{G}}^{(c)}_{s_{k_1} s_{k_2}  \sigma_{k_3} \sigma_{k_4}} }&=(\alpha_c -\bar \alpha_c)(z-\bar z)   \int {ds dt \over (4\pi i)^2}U^{{s + k^-_{43}\over 2}  } V^{{t - k_{14}^+ \over 2}  }\,  \widetilde{\Gamma}_{\{k_i \}}(s,t)  \sum_{m_1, m_2 =0}^{\infty}
\sigma ^{m_2 +{k^-_{21} + k^-_{43} \over 2}} \tau ^{k_1 -1 -m_{12}}  \cr
 &  \times {1 \over \prod_{i=1}^6 m_i!  }   \frac{ 2 (k^+_{24} - \tilde{u} )  (k^+_{13}-\tilde{u}  )   }{3 ( s+2m_1
 -{k_{12}^+ }+2)  
  ( {t}  + {k^-_{14} } -2 m_{12} -2) (-{\tilde{u} } + k_{13}^+  -2m_2 )} \, , 
     \end{align} 
where
\begin{align}
m_3 &=  k_1 - m_{12} -2\, , \qquad m_4 =  {k^-_{31} + k^-_{42} \over 2}   + m_1 \, , \cr
m_5 &=  {k_{12}^+ + k^-_{34} \over 2}  - m_{12} -2 \, , \qquad m_6 ={k^-_{21} +k^-_{43} \over 2} +m_2\, . 
  \end{align}

The full non-reduced correlator is obtained by putting back the factor given in \eqref{eq:R-factor}. To have a well-defined non-reduced Mellin amplitude accodring to \eqref{eq:Mellin2}, one may combine the factor $(\alpha_c -\bar \alpha_c)(z-\bar z)$ in \eqref{eq:chiral} with the same factor in \eqref{eq:R-factor}, so that $(\alpha_c -\bar \alpha_c)^2(z-\bar z)^2$ is a simple polynomial in $U, V$ and $\sigma, \tau$, given as
\ie \label{eq:zzb}
(\alpha_c -\bar \alpha_c)^2 (z-\bar z)^2 = \tau^{-2} \left[ (\sigma-\tau)^2 -2 (\sigma + \tau) +1 \right] \! \left[ (U- V)^2-2 (U+ V)+1 \right]\, .
\fe
So for this particular part of the chiral sector contribution, the non-reduced Mellin amplitude is well-defined, and can be obtained explicitly from \eqref{eq:zzb} and using the relation \eqref{eq:MtM}. We find that, interestingly, this term vanishes in the flat-space limit. More precisely, the leading two-derivative contribution arising from each term in \eqref{eq:zzb} cancels out. We also note ${\widetilde{\mathcal{G}}^{(c)}_{s_{k_1} s_{k_2}  \sigma_{k_3} \sigma_{k_4}}}$ vanishes in the MRV limit, due to $(\alpha_c -\bar \alpha_c)=0$ in the limit.  

Another important feature is that the multiple poles in \eqref{eq:chiral} do not cancel out even after converted into non-reduced Mellin amplitude using \eqref{eq:zzb}, as described above. As we will come back to this in the next section, these multiple poles precisely cancel with the same multiple poles arising from the non-chiral sector ${\widetilde{\mathcal{G}}^{(nc)}_{s_{k_1} s_{k_2}  \sigma_{k_3} \sigma_{k_4}}}$, such that the full non-reduced Mellin amplitude only contains single poles, as it should be for a local theory. 

\subsubsection{The non-chiral sector} 

As shown in \eqref{eq:Gnc}, the recursion relation for the non-chiral sector $\widetilde{\mathcal{G}}^{(nc)}_{s_{k_1} s_{k_2}  \sigma_{k_3} \sigma_{k_4}}$ is clearly more complicated, which however is a function of $U, V$ only and has a well-defined Mellin representation.  We have performed the contour integrals following the same methods, however the answer obtained in this way turns out to be rather lengthy, and it is not illuminating to present the expression here. Roughly, each term in \eqref{eq:Gnc} gives an expression that is similar to that of the correlators in tensor multiplet as given in \eqref{eq:non-red} or those of the chiral sector as given in \eqref{eq:chiral}, and it is not clear how to combine these terms together and simplify them.  However, we find that an equivalent but much more compact expression can be obtained by exploring the analytic structures of the correlators in Mellin space. That is what we will present in the following.

In particular, we express the result to manifest the pole structures of the Mellin amplitude, 
\ie \label{eq:gravitytensor}
\widetilde{\mathcal{M}}^{(nc)}_{s_{k_1} s_{k_2}  \sigma_{k_3} \sigma_{k_4}} (s,t)= \widetilde{\mathcal{M}}^{(nc)}_s(s,t) +  \widetilde{\mathcal{M}}^{(nc)}_{s,t}(s,t) +  \widetilde{\mathcal{M}}^{(nc)}_{s,u}(s,t)\,, 
\fe
where $\widetilde{\mathcal{M}}^{(nc)}_s(s,t)$ represents terms with single poles, $\widetilde{\mathcal{M}}^{(nc)}_{s,t}(s,t)$ is the contribution that has simultaneous poles in both $s$ and $t$ channels and similarly $\widetilde{\mathcal{M}}^{(nc)}_{s,u}(s,t)$ contains $s$- and $u$-channel poles. Furthermore, $\widetilde{\mathcal{M}}^{(nc)}_{s,t}(s,t)$ and $\widetilde{\mathcal{M}}^{(nc)}_{s,u}(s,t)$ are related to each other by a simple permutation,
\begin{align} \label{eq:su}
\widetilde{\mathcal{M}}^{(nc)}_{s,u}(s,t)=\widetilde{\mathcal{M}}^{(nc)}_{s,t}(s,\tilde{u}) \left.\right|_{k_1\leftrightarrow k_2,\, \s \leftrightarrow  \tau} \, .
\end{align}
Therefore, we will focus on $\widetilde{\mathcal{M}}^{(nc)}_{s,t}(s, t)$ only.

Let us begin with the single-pole term, $\widetilde{\mathcal{M}}^{(nc)}_s(s,t)$.  We find this term only contains $s$-channel poles, and according to its behaviour as polynomials in $\sigma$ and $\tau$, we find it is convenient to write $\widetilde{\mathcal{M}}^{(nc)}_s(s,t)$ as, 
\begin{align} \label{eq:Mnc}
\widetilde{\mathcal{M}}^{(nc)}_s(s,t)=\sum_{s_p=0}^{s_{\rm max}}  \sum_{j=j_{\rm min}}^{j_{\rm max}}\frac{\mathcal{R}_{s_p,1}^{j} \tau+ \mathcal{R}_{s_p,\,0}^{j}  }{s-s_p}   \s^j  \tau^{\frac{s_p+k^-_{43}}{2} -j-1}  \, ,
\end{align}
where $s_{\rm max}= {\rm min}\{k_{12}^+,k_{34}^+\}-2$, and the $j$-sum runs from $j_{\rm min}={\rm max} \{0,\frac{k_{12}^-+k_{34}^-}{2}\}$ to $j_{\rm max}={\rm min} \{\frac{s_p-k_{12}^-}{2},\frac{s_p-k_{34}^-}{2}\}$. The residues $\mathcal{R}_{s_p,1}^{j}, \mathcal{R}_{s_p,\,0}^{j}$ are independent of $\s$ and $\tau$, and they are given by
\ie \label{eq:RR}
\mathcal{R}_{s_p, 1}^{j} &= -\frac{ (-1)^{\frac{k_{1234}^+}{2} +k^-_{34}} \left(k_3^2+k_4^2-s_p (s_p+2)-2\right) }{3\, \Gamma^j_{\otimes}} \, , \\
\mathcal{R}_{s_p,\,0}^{j} &=- \frac{(-1)^{\frac{k_{1234}^+}{2} + k^-_{34}} \left(s_p-k^-_{12}-2j\right) 
\left(s_p-k^-_{34}-2j\right)}{3\, \Gamma^j_{\otimes}} \, ,  
\fe
where $\Gamma^j_{\otimes}$ is a product of $\Gamma$ functions
\begin{align} \label{eq:otimes}
\Gamma^j_{\otimes}= \;
&\Gamma (j+1) \,\Gamma \left(\frac{k^-_{12}+k^-_{34}}{2} +j+1 \right)  
 \Gamma \left(\frac{k_{12}^+-s_p}{2} \right) \Gamma \left(\frac{k_{34}^+-s_p}{2}  \right)\nonumber \\
& \times  \Gamma \left(\frac{k^-_{21} +s_p}{2} + 1-j \right) \Gamma \left( \frac{k^-_{43} +s_p}{2} + 1-j \right)\,.
\end{align}
The term $\widetilde{\mathcal{M}}^{(nc)}_{s,t}(s,t)$, that contains poles in both $s$ and $t$ channels, has similar structures and takes the following form, 
\ie \label{eq:double_pole}
\widetilde{\mathcal{M}}^{(nc)}_{s,t}(s,t)=\sum_{s_p=0}^{s_{\rm max}}\sum_{ j=j_{\rm min}}^{j_{\rm max}}\frac{\mathcal{R}_{s_p, t_p}^{j}}{(s-s_p)(t-t_p )}  \, \sigma ^{j} \tau^{\frac{s_p + k^-_{43} }{2}-j-1}     \, , 
\fe
where $t_p=k_{13}^+-s_p +2j$, and we find the residue is given by 
\begin{align} \label{eq:double_pole_R}
\mathcal{R}_{s_p, t_p}^{j} &=
\frac{\, (-1)^{\frac{k^+_{1234}}{2}}}{3\, \Gamma^j_\otimes} \left(s_p+k^-_{21}-2j\right)
\left(s_p+k^-_{43}-2j\right)  \Big[ j \left(j-\frac{{k^-_{21}}+k^-_{43}}{2} \right) {x^2 \over \s} \\
&+ \left( j ({k^-_{21}}+{k^-_{43}}-2)+k_{24}^+-{s_p}-2-2 j^2\right)   x +  (2 j+1) ({k^-_{21}}+{k^-_{43}}-2) -4 j^2   \nonumber 
\\
&+\left({s_p} (3- k_{1234}^+ +s_p)+ ( k^+_{12}\,k^+_{34}- k^+_{13}-2 k_{24}^+ +4 ) -\frac{j}{2}  ({k^-_{21}}+{k^-_{43}}-4-2j)\right) \sigma \nonumber\\ &-\frac{1}{4}(k_{12}^+-{s_p}-2) (k_{34}^+-{s_p}-2)   (\sigma -x)^2\Big]\, , \nonumber
\end{align}
with $x=\tau-1$. Together with the result of $\widetilde{\mathcal{G}}^{(c)}_{s_{k_1} s_{k_2} \sigma_{k_3} \sigma_{k_4} }$ in \eqref{eq:chiral}, we obtain the complete solution for the holographic correlator $\langle s_{k_1}s_{k_2}\s_{k_3} \s_{k_4}\rangle$. 

A few comments are in order. Firstly, we have verified that the simplified expression \eqref{eq:gravitytensor} agrees with the result obtained directly from solving recursion relation using the methods similar to that in the previous section for studying the correlators of operators in tensor multiplet. Secondly, the compact expression we obtained here suggests that the holographic correlators in $AdS_3 \times S^3$ exhibit new structures that cannot be seen from hidden conformal symmetries, especially when operators in gravity multiplet are involved. As we have emphasised, the solution obtained directly from \eqref{eq:Gnc} is rather complex. The expression we presented in \eqref{eq:Mnc}  has quite different structure compared to the result \eqref{eq:non-red} for the correlators in tensor multiplet that is obtained directly from recursion relations. The compact expression \eqref{eq:Mnc} rather  has structures that are analogous to those of the references \cite{Alday:2020dtb, Alday:2020lbp}.  Finally, use the relation \eqref{eq:non-red2}, we can again obtain the non-reduced Mellin amplitude for the contribution of the non-chiral sector, $\mathcal{M}^{(nc)}_{s_{k_1} s_{k_2}  \sigma_{k_3} \sigma_{k_4}}(s,t )$.  Importantly, as we commented earlier, we find $\mathcal{M}^{(nc)}_{s_{k_1} s_{k_2}  \sigma_{k_3} \sigma_{k_4}}(s,t )$ contains multiple poles (arising from $\widetilde{\mathcal{M}}^{(nc)}_{s,t}(s,t)$ and $\widetilde{\mathcal{M}}^{(nc)}_{s,u}(s,t)$), however, these poles cancel precisely with those from the chiral sector. This cancellation provides a very non-trivial check on our results.

\subsection{The flat-space and MRV limits}

We will now study the flat-space and MRV limits of the non-reduced Mellin amplitude $\mathcal{M}_{s_{k_1} s_{k_2}  \sigma_{k_3} \sigma_{k_4}}(s,t )$. As we commented in the section \ref{sec:chiral}, the chiral sector contribution $\mathcal{G}^{(c)}_{s_{k_1} s_{k_2}  \sigma_{k_3} \sigma_{k_4}}$ does not contribute in these limits. So we will focus on the contribution from the non-chiral sector. 

Let us begin with flat-space limit. It is easy to see that the Mellin amplitude is dominated by the single pole term $\widetilde{\mathcal{M}}^{(nc)}_s(s,t )$ in this limit. The explicit form of the correlator in flat-space limit is given by
\begin{align} \label{eq:flat2}
\mathcal{M}^{(nc)}_{s_{k_1} s_{k_2}  \sigma_{k_3} \sigma_{k_4}}(\,s, t ) |_{s,t\rightarrow \infty} \rightarrow \frac{u\,t+s\,t\,\s+s\,u\,\tau}{s}\,P^{(nc)}_{\{k_i\}}(\s,\tau)   \, ,
\end{align}
where the overall factor $P^{(nc)}_{\{k_i\}}(\s,\tau)$ is a polynomial in $\s,\tau$, 
\begin{equation}
P^{(nc)}_{\{k_i\}}(\s,\tau) = -{1\over 4} \sum_{s_p=0}^{s_{\rm max}}\sum_{j=j_{\rm min}}^{j_{\rm max}} (\mathcal{R}_{s_p,1}^{j} \tau +\mathcal{R}_{s_p, 0}^{j} )\, \s^j \tau^{\frac{s_p+k^-_{43}}{2}-j-1}    \, ,
\end{equation}
with $\mathcal{R}_{s_p,1}^{j}$ and $\mathcal{R}_{s_p, 0}^{j}$ given in \eqref{eq:RR}.  We note, as required, the flat-space limit of the Mellin amplitude has a two-derivative power counting and has precisely the same structure as the flat-space superamplitude when compactified to 3D, as given in \eqref{eq:flat1}. In particular, they both only contain a single pole in $s$-channel. As in the case of tensor multiplet, we see again the appearance of the factor $(u\,t+s\,t\,\s+s\,u\,\tau)$, which represents the fact that the theory has half maximal supersymmetry.  We note for holographic correlators in other $AdS$ backgrounds, the corresponding flat-space prefactors that are analogous to $P^{(nc)}_{\{k_i\}}(\s,\tau)$ were derived as an overlap factor of the in- and out-states, which are dressed with nontrivial wavefunctions on a transverse sphere \cite{Alday:2021odx}. It would be very interesting to generalise the derivation for the correlators we consider here. 

The MRV limit of the non-reduced Mellin amplitude is defined as 
 \ie
{\mathcal{M}}_{{\rm MRV}, s_{k_1} s_{k_2}  \sigma_{k_3} \sigma_{k_4}}(s,t)={\mathcal{M}}^{(nc)}_{s_{k_1} s_{k_2}  \sigma_{k_3} \sigma_{k_4}}(s,t) \big{\vert}_{\alpha_c \rightarrow 0, \bar{\alpha}_c \rightarrow 0} \, .
\fe
We find that the term with $s, u$-channel  poles, $  \widetilde{\mathcal{M}}^{(nc)}_{s,u}(s,t)$, vanishes identically in the MRV limit, and the single-pole term, $  \widetilde{\mathcal{M}}^{(nc)}_{s}(s,t)$,  and the term with poles in $s, t$-channels, $\widetilde{\mathcal{M}}^{(nc)}_{s,t}(s,t)$, reduce to, 
\ie
{\mathcal{M}}_{{\rm MRV}, s_{k_1} s_{k_2}  \sigma_{k_3} \sigma_{k_4}}(s,t)= {\mathcal{M}}^{(nc)}_{{\rm MRV}, s}(s,t)+ {\mathcal{M}}^{(nc)}_{{\rm MRV}, s,t}(s,t)\,,
%+ {\mathcal{M}}^{(nc)}_{{\rm MRV}, s,u}(s,t)
\fe
where
\begin{align} \label{eq:MHV1}
{\mathcal{M}}^{(nc)}_{{\rm MRV}, s}(s,t) &=  (s+t-k_{13}^+)(s+t-k_{24}^+) \nonumber\\
  &\times \sum_{s_p=0}^{s_{\rm max}} \frac{1}{(s-s_p)}\frac{(-1)^{\frac{k_{1234}^+}{2} +k^-_{43}} \left(2-k_{21}^-\,k_{43}^--k_3^2-k_4^2- (k_{21}^-+k_{43}^- - 2)s_p\right)
}{12 \, \Gamma_{\otimes}^{j=0}}\,, \end{align}
and
\begin{align} \label{eq:MHV2}
{\mathcal{M}}^{(nc)}_{{\rm MRV}, s, t}(s,t)  &=
 (s+t-k_{13}^+)(s+t-k_{24}^+) \nonumber\\
   &\times \sum_{s_p=0}^{s_{\rm max}} \frac{1}{(s-s_p)\left(t-t_p\right)}
  \frac{(-1)^{\frac{k_{1234}^+}{2} + k^-_{43}} ({k^-_{21}}+{k^-_{43}}-2)(k_{21}^- + s_p)(k_{43}^- +s_p)
}{12\, \Gamma^{j=0}_{\otimes}  }\, ,
\end{align}
and $\Gamma^{j=0}_{\otimes}$ is given in \eqref{eq:otimes} with $j$ being set to $0$. Importantly, the apparent double poles in ${\mathcal{M}}^{(nc)}_{{\rm MRV}, s, t}(s,t)$ in fact cancel out after the sum.  This can be understood by the fact that the residues at the double poles are all proportional to $(\alpha-\bar{\alpha})^2$ (so that they cancel with the contributions from chiral sector), which vanishes identically in the MRV limit. 

We note that, there are no $u$-channel singularities since ${\mathcal{M}}^{(nc)}_{{\rm MRV}, s,u}(s,t)$ vanishes identically in the MRV limit as we commented earlier. Furthermore, the prefactor $(s+t-k_{13}^+)(s+t-k_{24}^+)$ in \eqref{eq:MHV1} and \eqref{eq:MHV2} gives arise zeros in $u$-channel. As we have already emphasised in the previous section when we studied the correlators of operators in tensor multiplet, these properties of the correlators in the MRV limit are crucial in the study of holographic correlators in other $AdS$ backgrounds.  

Finally, let us remark that one may consider the other MRV limit with $\alpha_c = \bar{\alpha}_c=1$, for which, we find that the chiral sector contribution $\mathcal{G}^{(c)}_{s_{k_1} s_{k_2}  \sigma_{k_3} \sigma_{k_4}}$ also vanishes due to the fact it is proportional to $\alpha_c - \bar{\alpha}_c$. The non-trivial contribution arising from the non-chiral sector is given by the terms that have leading order term in $\s, \tau$. Explicitly, we find 
\begin{align}
{\mathcal{M}}_{{\rm MRV'}, s_{k_1} s_{k_2}  \sigma_{k_3} \sigma_{k_4}}(s,t)= {\mathcal{M}}^{(nc)}_{{\rm MRV'}, s}(s,t)+ {\mathcal{M}}^{(nc)}_{{\rm MRV'}, s,t}(s,t)+ {\mathcal{M}}^{(nc)}_{{\rm MRV'}, s,u}(s,t)\,,
%+ {\mathcal{M}}^{(nc)}_{{\rm MRV}, s,u}(s,t)
\end{align}
where
\begin{align}  \label{eq:MHV1_prime}
{\mathcal{M}}^{(nc)}_{{\rm MRV'}, s}(s,t) &=  (k_{12}^+ -s)(k_{34}^+ -s)\,\s^{\frac{s_{\max}+k^-_{43}}{2}+1} \nonumber\\
  &\times  \sum_{j=j_{\rm min}}^{j_{\rm max}} \frac{1}{(s-s_{p}-2)}\left.\frac{\mathcal{R}_{s_{p},\,1}^{j}}{4\, \Gamma_{\otimes}}\right\vert_{s_p=s_{\rm max}}\,, \end{align}
  and 
\begin{align}  \label{eq:MHV2_prime}
{\mathcal{M}}^{(nc)}_{{\rm MRV'}, s,t}(s,t) &=  (k_{12}^+ -s)(k_{34}^+ -s)\,\s^{\frac{s_{\rm max}+k^-_{43}}{2}+1} \nonumber\\
  &\times \sum_{j=j_{\rm min}}^{j_{\rm max}} \frac{1}{(s-s_{p}-2)(t-t_p)}\left.\frac{\mathcal{R}_{s_p, t_p}^{j}}{4\, \Gamma_{\otimes}}\right|_{s_p=s_{\rm max}}\,, \end{align} 
and ${\mathcal{M}}^{(nc)}_{{\rm MRV'}, s,u}(s,t)={\mathcal{M}}^{(nc)}_{{\rm MRV'}, s,t}(s,u)|_{k_1 \leftrightarrow k_2}$. The apparent $s$-channel poles in \eqref{eq:MHV1_prime} and \eqref{eq:MHV2_prime} are always cancelled by the pre-factor $(k_{12}^+ -s)(k_{34}^+ -s)$, so there are no singularities in the $s$-channel. 
%%%%%
%%%%%
\section{Conclusion}
\label{secconclusion}
%%%%%

In this paper, we present compact formulas for all four-point tree-level holographic correlators in $AdS_3 \times S^3$ in supergravity limit, with all the operators in tensor multiplet, as well as for the mixed correlators where we have two operators in tensor multiplet and the other two in gravity multiplet. The formulas are obtained by solving recursion relations arising from a hidden 6D conformal symmetry of the theory \cite{Giusto:2020neo} that relates correlators of operators with higher weights to correlators of operators with lower weights. The recursion relation for the mixed correlators involving operators in gravity multiplet is relatively more complex compared to the one for the correlators involving only tensors. As we emphasised that the expression of the mixed correlators obtained directly from the recursion relation is rather lengthy, and a compact formula was found only after we carefully analyse the analytical properties of the correlators and re-express the result in a form that manifests the pole structures. The simple expression suggests new properties beyond the hidden conformal symmetry. It is therefore of interest to investigate if the expression can be obtained by other means. We also studied the structures of the correlators by taking various limits (that include flat-space limit and MRV limit) of the results, and interesting properties were found in these limits. We have further verified that the multiple poles cancel out non-trivially for the non-reduced Mellin amplitude. 

It will be of interest to extend the analysis to the correlators of four operators all in gravity multiplet, namely $\langle \sigma_{k_1} \sigma_{k_2} \sigma_{k_3} \sigma_{k_4}\rangle$. With the result of all these correlators, we will in principle complete the computation of all the four-point tree-level holographic correlators in $AdS_3 \times S^3$ in supergravity limit. It is expected that the correlators $\langle \sigma_{k_1} \sigma_{k_2} \sigma_{k_3} \sigma_{k_4}\rangle$ are described by a single 6D CFT correlator of four self-dual $3$-forms~\cite{Giusto:2020neo}, due to the conjectured hidden 6D conformal symmetry. The tree-level four-point correlators would allow the computation of CFT datas such as anomalous dimensions of non-BPS operators, some of which has been studied recently utilising the results of correlators in tensor multiplet \cite{Aprile:2021mvq, Ceplak:2021wzz}. The complete tree-level results would also allow the construction of loop corrections using analytical conformal bootstrap and unitarity methods.   The loop corrections for amplitudes in 6D $(2,0)$ supergravity are of particular interest, since the theory is anomalous only if we have the right matter content. The study of the anomaly in flat-space amplitudes in 6D $(2,0)$ supergravity was explored in \cite{Chen:2014eva}. It will be very interesting to extend these ideas to the holographic correlators in $AdS_3 \times S^3$. Finally, four-point correlators with special multiple particle operators in tensor multiplet have been recently studied in \cite{Ceplak:2021wzz}, and interesting structures were found, it is interesting to study analogous correlators but now involving operators in gravity multiplet.

%%%%% 
 \section*{Acknowledgements}
 %%%%%
 
 We would like to thank  Stefano Giusto, Rodolfo Russo, and  Alexander Tyukov for collaborations on related topics and useful conversations. We would also like to thank Marcel Hughes for helpful discussions and Rodolfo Russo for comments on the draft. The work is supported by the Royal Society grant RGF\textbackslash R1\textbackslash 180037. CW is supported by a Royal Society University Research Fellowship No. UF160350.

% \nocite{*}
	\bibliographystyle{ssg}
	\bibliography{AdS3}

\end{document}